\documentclass[12pt]{iopart}
\usepackage[latin2]{inputenc}
\usepackage{iopams,color}

\usepackage{verbatim}

\usepackage{graphicx}
\expandafter\let\csname equation*\endcsname\relax
\expandafter\let\csname endequation*\endcsname\relax
\usepackage{amsmath}
\usepackage{amssymb}
\usepackage{epstopdf}
\usepackage{hyperref}

\newcommand{\beq}[1]{\begin{equation} #1 \end{equation}}
\newcommand{\vc}[1]{\mathbf{#1}}
\newcommand{\expv}[1]{\left\langle #1 \right\rangle}

\begin{document}
\title{Effect of assisted hopping on thermopower in an interacting quantum dot}

\author{S. B. Tooski,$^{1,2}$ A. Ram\v sak$^{2,3}$, B. R. Bu{\l}ka$^1$, and R. \v{Z}itko$^{2,3}$}
\address{$^1$Institute of Molecular Physics, Polish Academy of Sciences, ul. M. Smoluchowskiego 17, 60-179 Pozna{\'n}, Poland}
\address{$^2$Jo\v zef Stefan Institute, Ljubljana, Slovenia}
\address{$^3$Faculty of Mathematics and Physics, University of Ljubljana, Ljubljana, Slovenia}
\ead{anton.ramsak@fmf.uni-lj.si}
\date{\today}

\begin{abstract}
We investigate the electrical conductance and thermopower of a quantum dot tunnel coupled to external leads described by an extension of the Anderson impurity model which takes into account the assisted hopping processes, {\it {\it i.e.}}, the occupancy-dependence of the tunneling amplitudes. We provide analytical understanding based on scaling arguments and the Schrieffer-Wolff transformation, corroborated by detailed numerical calculations using the numerical renormalization group (NRG) method. The assisted hopping modifies the coupling to the two-particle state, which shifts the Kondo exchange coupling constant and exponentially reduces or enhances the Kondo temperature, breaks the particle-hole symmetry, and strongly affects the thermopower. We discuss the gate-voltage and temperature dependence of the transport properties in various regimes. For a particular value of the assisted hopping parameter we find peculiar discontinuous behaviour in the mixed-valence regime. Near this value, we find very high Seebeck coefficient. We show that, quite generally, the thermopower is a highly sensitive probe of assisted hopping and Kondo correlations.
\end{abstract}

\submitto{\NJP}
\maketitle
\tableofcontents

\section{Introduction}
\label{intro}

The thermoelectric effect is the conversion of temperature differences to electric voltage and vice-versa. Thermoelectric devices find application in power generation, refrigeration, and temperature measurement \cite{Mahan}. The progress in nanotechnology has led to lower thermal conductivity while retaining the electrical conductivity and Seebeck coefficient \cite{Venkatasubramanian,Harman,Terasaki,Dubi}, which is
important for applied use.
In basic research, the thermoelectric effect is a tool for revealing the transport mechanisms. For instance, the position of the molecular states relative to the Fermi level can be deduced from the thermoelectric potential of molecular junctions \cite{Segal}.

Thermoelectric properties of nanomaterials are intensively studied \cite{Blanter,Kubala,Zianni,Rejec1,Sabzyan,Izadi,Wysokinski,Karamitaheri}. In transport through Coulomb islands some novel effects have been observed: the oscillations of the thermopower \cite{Blanter} and thermal conductance \cite{Kubala, Zianni} with gate voltage. In the coherent regime, the transport properties of quantum dots (QDs) attached to external leads strongly depend on the correlated many-body Kondo state. Its most notorious   signature is the increased conductance at low temperatures. Recently, there has also been growing interest in the thermopower of Kondo correlated quantum dots, which has been measured \cite{Godijn,Scheibner} and theoretically analyzed \cite{Costi,Andergassen,Schops,Rejec,Rejec2,Swirkowicz,Chen,Nguyen,Kim,Cornaglia,Tagani,Azema,Trocha,Kim2,Dong}.

Interacting QDs are commonly modelled using the single-impurity Anderson model (SIAM).
The most prominent term in this Hamiltonian is the on-site Coulomb repulsion.
The assisted-hopping terms arise as the next-leading effect of the Coulomb interaction after the on-site repulsion \cite{Hubbard,Dolcini,Amico,Dobry,Foglio,Simon1,Hirsch0,Simon2,Bulka,Rejec3,Hirsch1,Meir,Yu,Guinea,Hubsch,Hirsch2,Stauber,Borda,Lin,Penc,Kollar}.
Such processes are always present in real devices but are commonly neglected in theoretical modelling, despite the fact that they may, in fact, be quite sizeable.
A generalized Anderson impurity model with assisted hopping can be formally derived by integrating out high-energy degrees of freedom \cite{Foglio,Simon1}, leading to assised hopping to the retained level in the restricted basis. Such Hamiltonian can
also be postulated as a phenomenological model. The assisted-hopping terms first appeared in proposals for describing the properties of mixed-valence bulk systems. The discovery of high-$T_{c}$ superconductivity increased the interest in correlated hopping as a possible new mechanism for superconducting instability and temperature-induced metal-insulator transition \cite{Simon1,Hirsch0,Simon2, Bulka}. Furthermore, assisted hopping is considered as a significant factor in the stabilization of the ferromagnetism and localization of electrons \cite{Kollar}.

In the context of nanodevices, the assisted hopping has been proposed to account for anomalies in the conductance \cite{Meir} and  the unusual gate-voltage dependence of the measured Kondo temperature \cite{Yu} which does not follow the usual form expected for the standard SIAM. It has also been proposed to explain the conductance increase through local pairing effects \cite{Guinea}. It has been established that for transition-metal complexes the correlated hybridization can be very large, comparable even to the standard single-particle hybridization through the interatomic potential \cite{Hubsch}. For bulk systems, the inclusion of an assisted hopping term in the electronic Hamiltonian favors the existence of pairing correlations \cite{Hirsch2}. In the case of a QD, this tendency towards local pairing quenches the local moment, leads to asymmetries in the conductance of peaks associated with the same level \cite{Hirsch2}, and to a change of the thermopower in the sequential regime \cite{Wysokinski}.

In this work, we study the effect of the assisted-hopping term in a generalized SIAM on the  thermopower and conductance of the QD by applying the numerical renormalization group  (NRG) technique. We will show that the Kondo effect can be either suppressed or enhanced, depending on the gate voltage and the sign of the assisted-hopping term. This results from the renormalization of the level positions and widths, as well as the modification of the effective Kondo exchange coupling, which leads to exponential reduction of the Kondo temperature and consequently to a strong enhancement of the Seebeck coefficient. Another important effect is the particle-hole (p-h) symmetry breaking and the resulting asymmetry in the gate-voltage dependence of system properties.

\section{Model and method}
\label{model}

\subsection{Model}

We consider a QD described by an extended SIAM
\begin{eqnarray}
\label{qua1}
H&=& \sum_{\alpha, k,\sigma}\epsilon_{k} n_{\alpha k\sigma}
+\epsilon\; n+ U n_{\uparrow}n_{\downarrow}-V\sum_{\alpha, k,\sigma}[(1-x \,n_{\bar{\sigma}})c_{\alpha k \sigma}^\dag c_{\sigma}+{\rm H.c.}].
\end{eqnarray}
The first term corresponds to electrons in the leads (left and right, $\alpha=L,R$), $n_{\alpha k\sigma}=c_{\alpha k \sigma}^\dag c_{\alpha k\sigma}$ is the number operator for electron with wavevector $k$, spin $\sigma$, and energy $\epsilon_{k}=\langle k|h|k\rangle$; $h$ is the one-particle kinetic Hamiltonian. The second term describes electrons in the QD level with energy $\epsilon=\langle c|h|c\rangle$. The Coulomb repulsion energy for two electrons with opposite spin in the same level is $U=\langle cc|e^{2}/r|cc\rangle$. Here $n_{\sigma}=c_{\sigma}^\dag c_{\sigma}$ is the number operator for an electron in QD, and $n=n_{\uparrow}+n_{\downarrow}$. The last term describes the coupling between the leads and the QD, $V=-\langle k|h|c\rangle$ is the single-electron hopping parameter  between the impurity and the leads, which is positive for electron hopping. The assisted-hopping parameter $X=\langle kc|e^{2}/r|cc\rangle$ describes the Coulomb-interaction-mediated transfer of an electron from the state $|k\rangle$ in  the lead to the QD, when the QD is already occupied by an electron with the opposite spin $\bar{\sigma}$.

The parameter $X$ calculated for inter-atomic hopping in transition metals \cite{Hubbard,Kollar} and in copper oxides \cite{Gammel,Hirsch} is in the range $0.1-1$ eV. In semiconducting quantum dots Meir {\it et al.} \cite{Meir} estimated the ratio $X/V=0.65$ and showed that the assisted-hopping can lead to a significant reduction of the tunneling rate through the excited state. In the following, we use the normalized assisted-hopping parameter $x=X/V$ and we assume symmetric coupling to both leads and a flat density of states $\rho=1/(2D)$, where $D=1$ is the half-width of the conduction band. Since the value of $x$ can be of order $1$, the assisted hopping can significantly affect the transport properties. An interesting special case occurs for $x=1$, when the doubly-occupied state is fully decoupled from the leads and the transport is due solely to the singly-occupied states. The model is non-ergodic at $x=1$ and one expects anomalous properties.
Some vestigial effects are also expected for values of $x$ near 1. Negative values for $x$ are not excluded, {\it e.g.}, for the case of transport through molecules with higher angular momentum orbitals which can lead to positive or negative  hopping-overlap integrals.

\subsection{Perturbative analysis}

In the roughest mean-field treatment, the effective hybridization in this model is occupancy-dependent, $\Gamma_\mathrm{eff}=\Gamma (1-x \expv{n})^2$,
where $ \Gamma=2\pi \rho V^2 $.
This approximation is, however, overly simplistic. It is crucially important to account for the different rates of the $0 \leftrightarrow 1$ and $1 \leftrightarrow 2$ charge fluctuations. We expect two main effects: (i) different level renormalisations, and (ii) modified Kondo exchange coupling constant.

We discuss first the level renormalization in the spirit of the Haldane scaling where high
energy charge fluctuations are integrated out \cite{haldane1978,jefferson1977,martinek2005}. The zero-occupancy level $E_0$ is renormalized by the processes $0 \leftrightarrow \uparrow,\downarrow$. The singly-occupied level $E_1$ is renormalized by the processes $\sigma \leftrightarrow 0, 2$.
Finally, the doubly-occupied level is renormalized by the processes $2 \leftrightarrow \uparrow,\downarrow$. The renormalization of the lower atomic level $\epsilon_1=\epsilon$ can be extracted as the difference $E_1-E_0$, while the renormalization of the upper atomic level $\epsilon_2=\epsilon+U$ is obtained as $E_2-E_1$. The scaling calculations gives for $\epsilon_1$ a shift of
\beq{
\label{epsilon1}
\delta \epsilon_1 = -\frac{1}{\pi} \int d\omega \left\{
\frac{\Gamma [1-f(\omega)]}{\omega-\epsilon} + \frac{\Gamma (1-x)^2 f(\omega)}
{\epsilon+U-\omega} \right\},
}
while $\delta \epsilon_2=-\delta \epsilon_1$.
Here $f$ is the Fermi-Dirac distribution.
In the limit of standard SIAM, {\it {\it i.e.}}, for $x=0$, one can perform the integration exactly, the $T=0$ result being
\beq{
\delta \epsilon_1 = \frac{\Gamma}{\pi} \ln \frac{|\epsilon|}{|\epsilon+U|}.
}
It vanishes for $|\epsilon|=|\epsilon+U|$, {\it {\it i.e.}}, at the particle-hole
(p-h) symmetric point. In the presence of assisted hopping, the integration becomes more involved and cutoff dependent, thus no simple closed-form expression can be
provided. In any case, it is easy to see that the p-h transformation
with respect to the point $\epsilon=-U/2$ is no longer a symmetry of the system.

To find the Kondo exchange coupling constant $J_K$, we performed the Schrieffer-Wolff transformation, obtaining
\beq{
\label{JK1}
\rho J_K = \frac{2\Gamma(1-x)^2/\pi}{\epsilon+U}-\frac{2\Gamma/\pi}{\epsilon}.
}
At $\epsilon=-U/2$, this simplifies to $\rho J_K = \frac{4\Gamma}{\pi U} \left( 2-2x+x^2 \right)$, and for $x=0$ one recovers the standard result $\rho J_K=8\Gamma/\pi U$.
Equation~\eqref{JK1} succinctly shows the general trend:
at the p-h symmetric point of the standard SIAM, the Kondo coupling is renormalized by the assisted-hopping simply by a multiplicative
factor $(2-2x+x^2)/2$. This function is a parabola with a minimum at $x=1$, where it
has value $1/2$. Thus at
$\epsilon+U/2=0$, going from $x=0$ to $x=1$ the Kondo coupling will be reduced by half, resulting in an exponentially strong suppression of the Kondo temperature \cite{Hewson}
\beq{
T_K = c(U,\Gamma) \sqrt{\rho J_K} \exp\left(-\frac{1}{\rho J_K} \right)
}
where $c(U,\Gamma)$  is
the $U$ and $\Gamma$ dependent effective bandwidth.
For positive $x<1$ we thus expect a rapidly decreasing $T_K$ for increasing $x$. For $x>1$, $J_K$ starts to increase again,
and at $x=2$ it has the same value as for $x=0$. For $x<0$, the Kondo coupling
increases with the absolute value of $x$, thus $T_K$ is exponentially enhanced.

At other values of the gate voltage, the renormalization effects of the assisted
hopping need to be estimated from the more general equation~\eqref{JK1}. For $x=0$,
the Kondo coupling attains its minimum value at
$\epsilon=-U/2$. Minimizing the expression in equation~\eqref{JK1} we find that
the minimum is in general shifted to
\beq{
\epsilon = -\frac{U}{{1+|1-x|}},
}
{\it {\it i.e.}}, to smaller values of the gate voltage for increasing $x$. A highly curious feature
is that at $x=1$, we find $\epsilon=-U$, {\it {\it i.e.}}, the gate voltage that corresponds
to the edge of the Kondo plateau in the standard $x=0$ SIAM. At $x=1$, instead, we there find a minimum of $J_K$ and deep Kondo regime. For $x\to 1$, we will thus expect a spectacular asymmetry where the Kondo plateau extends right up to the point where the occupancy should suddenly go to 2, because the decoupled upper effective atomic level at $\epsilon+U$ falls below the Fermi level. Since the deep Kondo regime is associated with single occupancy, this observation
suggests anomalous behaviour for $x \to 1$ for gate voltage near $\epsilon=-U$.
This is indeed fully corroborated by the numerical results presented in the following.

The system possess an exact symmetry around $x=1$, which holds beyond the perturbative derivation of the Kondo coupling $J_K$. In particular,
all results presented in this paper exhibit a perfect symmetry with respect to the $x=1$ point, {\it {\it i.e.}}, the results calculated at $x$ are identical to those obtained at $2-x$.
This can be understood as follows. For $n_{\bar{\sigma}}=0$, the
hopping matrix element of spin-$\sigma$ electron is $-V$, while for
$n_{\bar{\sigma}}=1$ it equals $-V(1-x)$. Replacing $x$ with $2-x$,
for $n_{\bar{\sigma}}=0$ the matrix element remains unchanged, while
for $n_{\bar{\sigma}}=1$ we find $V(1-x)$, which differs only in sign.
The sign is, however, immaterial. Each electron hop in one direction
must be followed by another in the opposite direction in order to obtain a
state which is not orthogonal to the original state. The statistical
sum thus only depends on $x$ through the combination $V^2(1-x)^2$,
hence all system properties are symmetric with respect to $x=1$.

In numerical calculations presented in this work we set the hybridization strength to $\Gamma/D=0.02$ and we focus on the stronger coupling,
$U/\Gamma=8$, where the model enters the Kondo regime near half-filling; for $x=0$, $T_K$ is of order $10^{-3}$. For convenience, we fix the Fermi level at zero, $\epsilon_F=0$.

\subsection{Transport coefficients}

Thermoelectric transport is calculated for a situation in which a small external bias voltage, $\delta V = V_{L} - V_{R}$, and a small temperature gradient $\delta T$ are applied between left and right leads  \cite{Costi}. Left and right leads are then at different chemical potentials $\mu_{L}$ and $\mu_{R}$ and temperatures $T_{L}$ and $T_{R}$, with $e\delta V = \mu_{L} - \mu_{R}$ and $\delta T=T_{L}-T_{R}$. To  linear order, the following expressions for the electrical conductance, $G(T)$, and the thermopower (Seebeck coefficient), $S(T)$, are obtained
\begin{eqnarray}
\label{Gt}
G(T)=e^{2}I_{0}(T),
\end{eqnarray}
\begin{eqnarray}
\label{S}
S(T)=-\frac{|e|}{k_B T}\frac{I_{1}(T)}{I_{0}(T)},
\end{eqnarray}
where $I_{n}$ are the transport integrals
\begin{eqnarray}
\label{In}
I_{n}(T)=\frac{2}{h}\int d\omega\, \omega^{n}\, \mathcal{T} (\omega) \left(-\frac{\partial f}{\partial \omega}\right).
\end{eqnarray}
Here, $e$ denotes the unit charge and $h$ the Planck's constant.
The transmission coefficient $\mathcal{T}(\omega)$ is given by
$\mathcal{T}(\omega) = \pi \Gamma A(\omega)$ with $A(\omega)=-\frac{1}{\pi} \mathrm{Im} \langle\!\langle a_{\sigma};
a_{\sigma}^\dag \rangle\!\rangle_{\omega+i0^+}$,
with the operator $a_\sigma^\dagger = (1-x\,n_{\bar\sigma})c_\sigma^\dagger$. This is similar to the standard SIAM, but with a spectral function of the
operator $a_{\sigma}^\dagger$, rather than $c_\sigma^\dagger$, as can be seen
from the Dyson equation for the $G_{k,k'}$ Green's function in the leads.
It can namely be shown that the $T$-matrix is given by the correlator
of the $[H_\mathrm{hyb},c_{k\sigma}]$ objects, where $H_\mathrm{hyb}$ is the
hybridization part of the Hamiltonian (term proportional to $V$ in equation~\ref{qua1}).

To evaluate the transport integrals we used the numerical renormalization group (NRG) method \cite{Wilson,Krishna,Bulla}. This method allows to calculate static, dynamic and transport properties in a reliable and rather accurate way in a wide temperature range. The approach is based on the discretization of the continuum of bath states, transformation to a chain Hamiltonian, and iterative diagonalization. The calculations reported here have been carried out for a discretization parameter $\Lambda=2$, twist averaging over $N_{z}=4$ interleaved discretization meshes, and retaining 500 states per NRG step.

\begin{figure}[t]
\centering
\includegraphics[width=0.9\textwidth,clip]{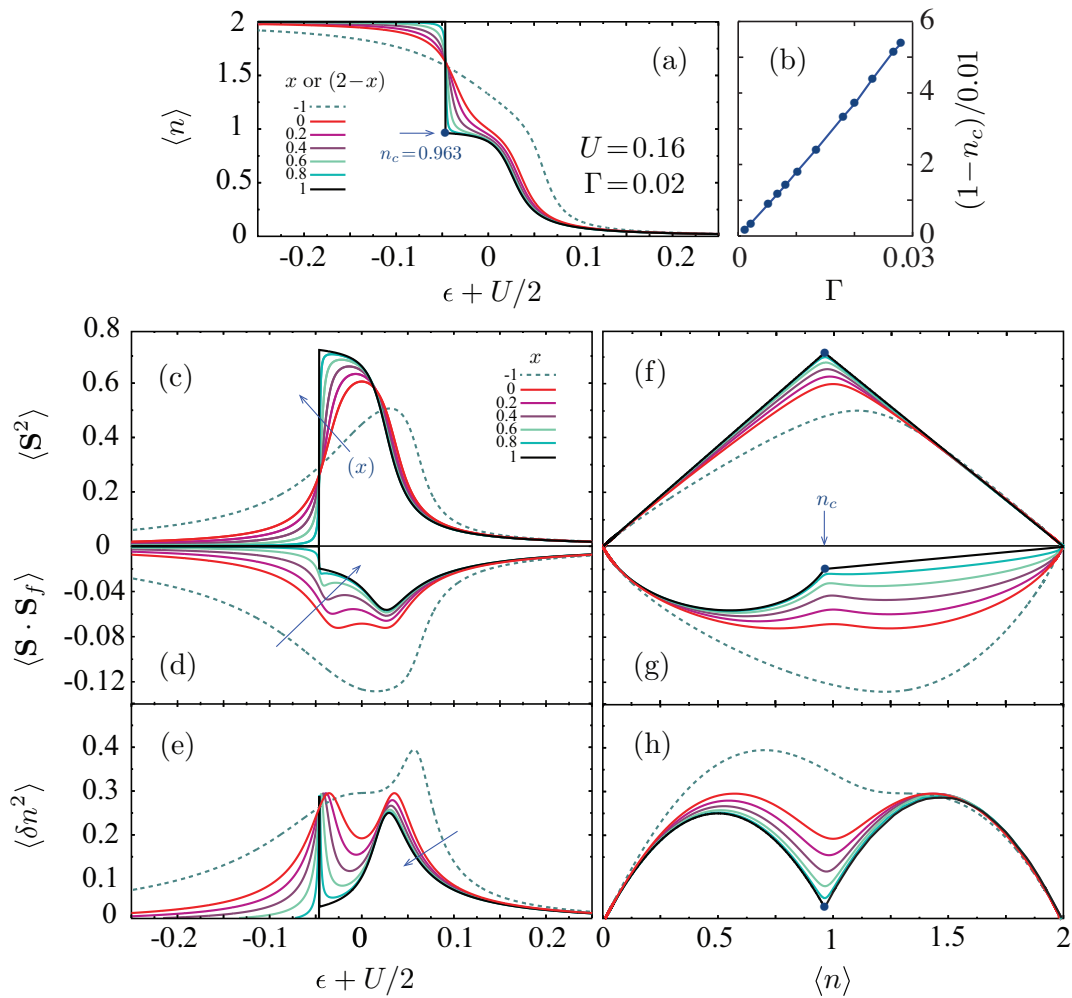}
        \caption{\label{fig1}
(a)  Total impurity charge (occupancy) $\langle n\rangle$ as a
function of gate voltage $\epsilon+U/2$ for various $x$ (and, due to the symmetry, also $2-x$). Parameters are $\Gamma=0.02$,  $U=0.16$, and $T=10^{-5}$. The arrow indicates critical $n_c$. (b) Critical $n_c$ as a function of the hybridization $\Gamma$. (c) Local spin squared (static local moment), (d) spin correlation between the QD level and the leads, and (e) local charge fluctuations, all as a function of the gate voltage (arrows indicate increasing $x$).  In panels (f), (g), and (h) the same quantities are shown as a function of the occupancy $\langle n\rangle$. Bullets indicate values at $\langle n \rangle =n_c$.}
\end{figure}

\subsection{Different regimes}

Near half-filling, the local-moment (LM) regime corresponds to the formation of a localized spin on the dot at intermediate temperatures ($T_{K} \lesssim T \lesssim U$). In this regime, the conductance has logarithmic temperature dependence which is the signature of the emerging Kondo state. At low
temperatures, $T \lesssim T_K$, {\it {\it i.e.}}, in the strong-coupling (SC) regime, the moment is fully screened and the system is characterized by Fermi-liquid properties.

The mixed-valence (MV) regime corresponds to gate voltages where the charge on the dot fluctuates between $0$ and $1$, or between $1$ and $2$. The physics is then governed by charge fluctuations and the system acts as a two noninteracting resonant-level model. Finally, in the empty-orbital (EO) regime, $n\approx 0$,
and the full-orbital (FO) regime, $n \approx 2$, there are only thermally activated charge fluctuations and the system behaves as a noninteracting resonant-level model.

\section{Static properties and spectral densities}

In figure~\ref{fig1}, the static quantities at low temperature $T=10^{-5}$ are plotted in left panels as a function of $\epsilon$ which is shifted by the gate voltage. The results for the standard SIAM, $x=0$, are well known: for $\epsilon+U/2 \approx 0$ the occupancy $\langle n\rangle$ tends to be pinned to 1 near half-filling, see panel (a); an even more pronounced plateau in $\langle n\rangle$ vs. $\epsilon$ curves would form in the large $U/\Gamma$ limit. In this gate-voltage range, the local moment is formed, thus $\expv{\vc{S}^2}$ is large, panel (c), and the antiferromagnetic exchange interaction between the dot level and the leads is signaled by negative
values of $\expv{\vc{S} \cdot \vc{S}_f}$, panel (d). Here $\vc{S}$ and $\vc{S}_f$ are the spin operators for the QD and the first site in the leads, respectively. The charge fluctuations $\expv{\delta n^2} = \expv{(n-\langle n\rangle)^2} = \expv{n^2}-\expv{n}^2$ have a (local) minimum near half-filling, but are enhanced in the MV regimes for $\epsilon \approx 0$ and $\epsilon+U \approx 0$, panel (e).

As expected on the basis of equation~\eqref{epsilon1}, the filling of the dot with electrons is significantly affected by the
assisted hopping.
At fixed gate voltage, the occupancy is increased for $\epsilon+U \lesssim 0$
and decreased for $\epsilon+U \gtrsim 0$ if $0<x<2$, and vice-versa for $x<0$ and $x>2$.
Another notable effect is the breaking of the p-h symmetry. This is a trivial consequence of
different hopping rates for $0 \leftrightarrow 1$ and $1 \leftrightarrow 2$ processes in the presence
of assisted hopping.

The assisted hopping in the range $0<x<2$ enhances the local moment in the gate voltage range $-U/2 \lesssim \epsilon+U/2 \lesssim 0$, as visible from the increased values of the
$\expv{\vc{S}^2}$ curves. This is a consequence of the reduced Kondo coupling
$J_K$, see equation~\eqref{JK1}, because it makes the local spin more decoupled from the leads. This is also mirrored in the decreasing absolute
value of $\expv{\vc{S} \cdot \vc{S}_f}$. For $\epsilon+U/2 \gtrsim 0$, the
local moment is reduced, but only very slightly. This small reduction can be thought of as a higher-order effect of the assisted hopping in the regime of small $\expv{n_{\bar\sigma}}$. The correlation
$\expv{\vc{S} \cdot \vc{S}_f}$ is also affected more mildly in this
gate-voltage range.

The charge fluctuations $\expv{\delta n^2}$ are particularly interesting, since they are directly
affected by the assisted-hopping term. For $0<x<1$ we observe that the
fluctuation peak at $\epsilon+U \sim 0$ becomes increasingly narrow with increasing $x$:
the parameter range of the valence-fluctuation region $1 \leftrightarrow 2$ is shrinking.
In fact, it becomes extremely small in the $x\to 1$ limit.
The width of the peak at $\epsilon \sim 0$ is, however, not significantly affected.

The particular behaviour in the occupancy can be understood analyzing the single particle Green's function derived within the Hubbard-I approximation \cite{Hubbard},
\begin{eqnarray}
\label{HI}
\langle\!\langle c_{\sigma};c_{\sigma}^\dag\rangle\!\rangle_{\omega} \simeq \frac{1-\langle n\rangle/2}{\omega-\epsilon-i \Gamma/2}+ \frac{\langle n\rangle/2}{\omega+-\epsilon-U-i(1-x)^2 \Gamma/2}\;.
\end{eqnarray}
This result is beyond the mean-field approximation and takes into account the Coulomb blockade, but neglects the spin-flip processes leading to the Kondo resonance (see also Ref.~\cite{Lacroix}).
The spectral function given by (\ref{HI}) has two peaks at $\omega=\epsilon$ and  $\omega=\epsilon+U$ with different width. The width of the excitation peak at $\omega= \epsilon+U$ shrinks to zero for $x=1$, becoming a delta peak, thus the occupancy jumps when the delta peak crosses $\epsilon_F$ as the gate voltage is swept,
from $\expv{n}=n_c \sim1$ to  exactly $\expv{n}=2$.
All other
quantities also exhibit  sharp transitions across this occupancy jump. In panel (a) the position of $\langle n \rangle=n_c$ near the transition is indicated by an arrow. Note that $n_c<1$, which can be explained as an effect of the charge fluctuations, since critical $1-n_c$  scales as $\Gamma/U$, as is clearly seen from panel (b).

In panels (f), (g), and (h) we show static quantities as a function of
the occupancy $\langle n \rangle $. This alternative representation reveals
additional effects of the assisted hopping beyond the dominant effect
({\it i.e.}, the modification of the filling dependence $\expv{n}$ vs.
$\epsilon$). We note a reduction of the charge fluctuations and an
increase of the moment which are rather symmetric with respect to
$\expv{n}=n_c\sim1$. This result can be fully accounted for within the simple
Hubbard-I approximation which properly describes the high-energy
charge fluctuations. The reduction of the $\expv{\vc{S}\cdot\vc{S}_f}$
correlations is, however, quite asymmetric. Within a simple
approximation, this expectation value is proportional to $J_K
\sqrt{\expv{\vc{S}^2}}$; this explains the overall shape of the curve,
and in particular the
asymmetry which is due to the asymmetry of $J_K$. Note also the $x=1$
results: $\langle {\bf S}^2\rangle=\frac{3}{4}\langle n \rangle$
for $\langle n \rangle<n_c$, because $\expv{n_\uparrow
n_\downarrow}=0$, {\it i.e.}, because the doubly-occupied state is fully
decoupled. Extremely interesting is also the $\langle \delta n^2
\rangle$ vs. $\langle n \rangle $ plot, panel (h). For $n_c< \langle n
\rangle <2$, the gate-voltage dependence is squeezed into a narrow
peak [panel (e)] in the $x\to1$ limit, yet the dependence on $\langle
n \rangle $ reveals similarity to results for $x<1$.
Note that the curve is continuous, because the calculation is
performed at finite temperature; strictly at $T=0$ there would be a true
discontinuity. Finally, we observe that the value at the minimum of the
$\langle \delta n^2 \rangle$ curve at $n_c$, indicated by the bullet
in panel (h), also scales with $\Gamma/U$ (scaling not shown here).

\begin{figure}[t]
\centering
\includegraphics[width=0.9\textwidth,clip]{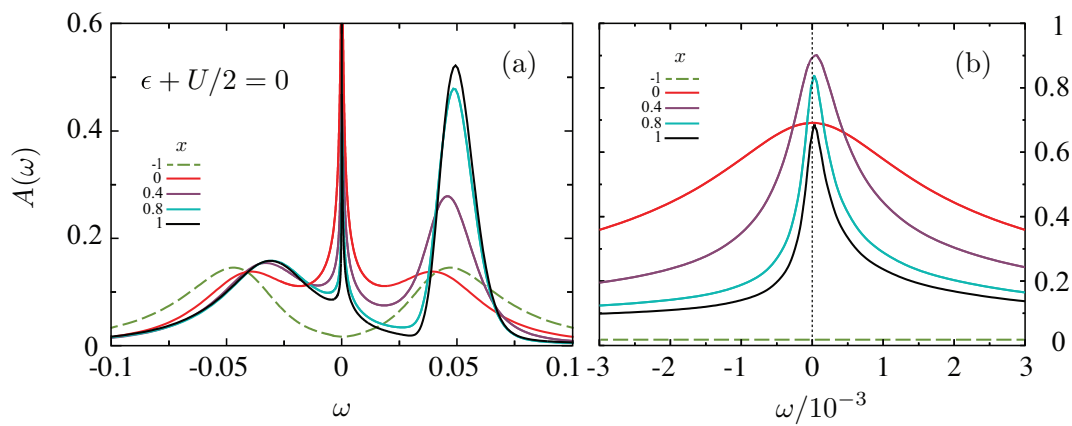}
        \caption{\label{fig2}
(a) Spectral function $A(\omega)$ for different values of the assisted hopping parameter $x$ at $\epsilon+U/2=0$ and $T=10^{-5}$. Note the absence of the Kondo peak for $x=-1$ (or for $x=3$). (b) Close-up of the Kondo peak region.}
\end{figure}

The generalized local spectral densities $A(\omega)$
in the Kondo regime (at $\epsilon+U/2=0$) are shown in figure~\ref{fig2} for various $x$.
At low temperatures, $T<T_{K}$, $A(\omega)$ is characterized by three peaks: the  Kondo resonance at $\epsilon_F$,
and two atomic resonances: lower peak at $\omega=\epsilon+\delta \epsilon_1$ and upper peak $\omega=\epsilon+U+\delta\epsilon_2$.

The narrow Kondo resonance for $0<x<2$ lies very near $\epsilon_F$, see panel (b). Its width is of order $T_{K}$ and its reduction for $0<x<1$ is manifest.
For $x>2$ (or, alternatively, for $x<0$), the Kondo resonance decreases rapidly in height and eventually disappears completely, see the dashed line in panel (b), since
the system moves away from the Kondo regime.

The atomic resonances contain most of the spectral weight and have widths approximately $2\Gamma$ and $2\Gamma(1-x)^2$.
Accordingly, the lower atomic peak does not depend much on $x$, while the other becomes
narrower with increasing $x$ in the range $0<x<1$.

\section{Thermopower and conductance}
\subsection{Gate-voltage dependence }

\begin{figure}[t]
\centering
\includegraphics[width=0.9\textwidth,clip]{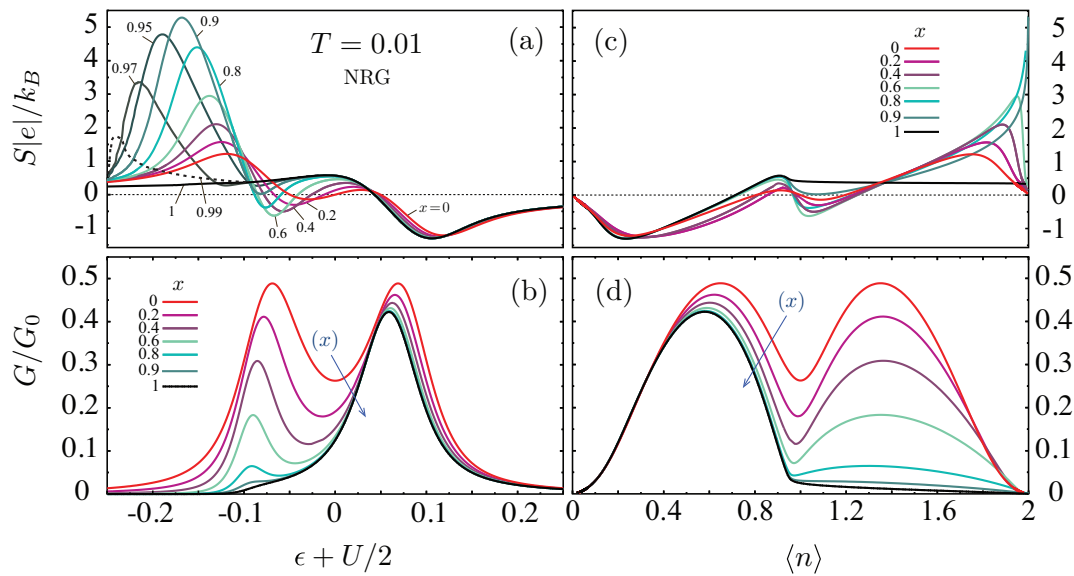}
\caption{\label{fig3}(a) Thermopower $S$ and (b) conductance $G$ results obtained by the NRG, as a function of the gate voltage  for different values of assisted hopping $x$, at $T/D=0.01$.
In panels (c) and (d) we show $S$ and $G$ as a function of $\langle n\rangle$.}
\end{figure}

We now study the influence of the assisted hopping on the transport properties. Thermopower and conductance are plotted in figure~\ref{fig3}(a) and (b), respectively,
as a function of the gate-voltage for constant high temperature $T=0.01$.
This temperature is above the Kondo temperature for all $0<x<2$.
$G$ is characterized by two conductance peaks associated with the levels $\epsilon$
and $\epsilon+U$, separated by the Coulomb blockade valley.
For $x=0$, the peaks are symmetric with respect to $\epsilon+U/2=0$ due to the p-h symmetry. $S$ is relatively high and its $\epsilon$-dependence is similar to that derived from the master equation for sequential tunneling through a quantum dot \cite{Beenakker}. Thermopower is positive for a large negative gate voltage, when transport is through holes. $S$ changes sign at $\epsilon+U \approx 0$,  when $G$ reaches its maximum. Next, $S$ goes through zero at the p-h symmetric point and becomes positive. At $\epsilon \approx 0$, $S$ changes sign yet again and becomes negative -- now the transport is dominated by electrons.

The assisted hopping modifies the spectral peaks at  $\epsilon$ and $\epsilon+U$, their widths being proportional to $\Gamma$ and $\Gamma(1-x)^2$, respectively. One may use the formula $S=\sum_n G_n S_n/\sum_n G_n$ for the thermopower in a system with several types of charge carriers $\{n\}$, each of which would separately have a contribution $S_n$ to thermopower and $G_n$ to conductance \cite{Mott}. In our model, we have two resonant channels with different tunneling rates $\Gamma$ and $\Gamma(1-x)^2$. The corresponding transmission function can be expressed as
\begin{eqnarray}
\mathcal{T}(\omega)= \sum_{n=1,2} \frac{z_n\Gamma^2_n/4}{(\omega-\epsilon_n)^2+\Gamma^2_n/4}\,
\label{res-trans}
\end{eqnarray}
where $\epsilon_1=\epsilon$,  $\epsilon_2=\epsilon+U$, $\Gamma_1=\Gamma$ and   $\Gamma_2=\Gamma (1-x)^2$. Here $z_1=1-\langle n\rangle/2$ and $z_2=\langle n\rangle/2$ are the weights of the resonant states. The same result can be derived within the Hubbard-I approximation, equation (\ref{HI}). Under these assumptions, the thermopower can be derived exactly (see Ref.~\cite{Rejec}),
\begin{eqnarray}
S=-\frac{k_B}{|e|}\;2\pi \dfrac{ \displaystyle\sum_n z_n\Gamma_n\;{\rm Im} \Bigl[\dfrac{\Gamma_n/2+i\epsilon_n}{2\pi k_B T}\; \psi'\Bigl(\dfrac{1}{2}+\dfrac{\Gamma_n/2+i\epsilon_n}{2\pi k_B T}\Bigr)\Bigr] } { \displaystyle\sum_n z_n\Gamma_n\; {\rm Re} \Bigl[\psi'\Bigl(\dfrac{1}{2}+\dfrac{\Gamma_n/2+i\epsilon_n}{2\pi k_B T}\Bigr)\Bigr] },\,
\label{S-res}
\end{eqnarray}
where $\psi'(z)=\sum_{n=0}^{\infty} (z+n)^{-2}$ is the tri-gamma function \cite{Abramovitz}. The numerical results for $x\neq0$, figure~\ref{fig3}(a), confirm the composition of contributions from both resonant transmissions. $S$ changes its sign at $\epsilon+U/2 \sim -U/2$ and at $\epsilon+U/2 \sim U/2$, when the atomic levels cross $\epsilon_F$. Since the assisted hopping breaks the p-h symmetry, the curves are asymmetric and shifted to lower $\epsilon$ values for $0<x<1$. One can see a large peak in $S$ for $x=0.9$ which is attributed to  the resonant transmission through the level at $\epsilon+U$.  At this temperature the corresponding thermopower $S_2$ is very large because the hopping transport dominates. For $x=1$ the peak disappears and the thermopower  is determined by the transport through the lower level at $\epsilon$ only, $S=S_1$.

\begin{figure}[t]
\centering
\includegraphics[width=0.9\textwidth,clip]{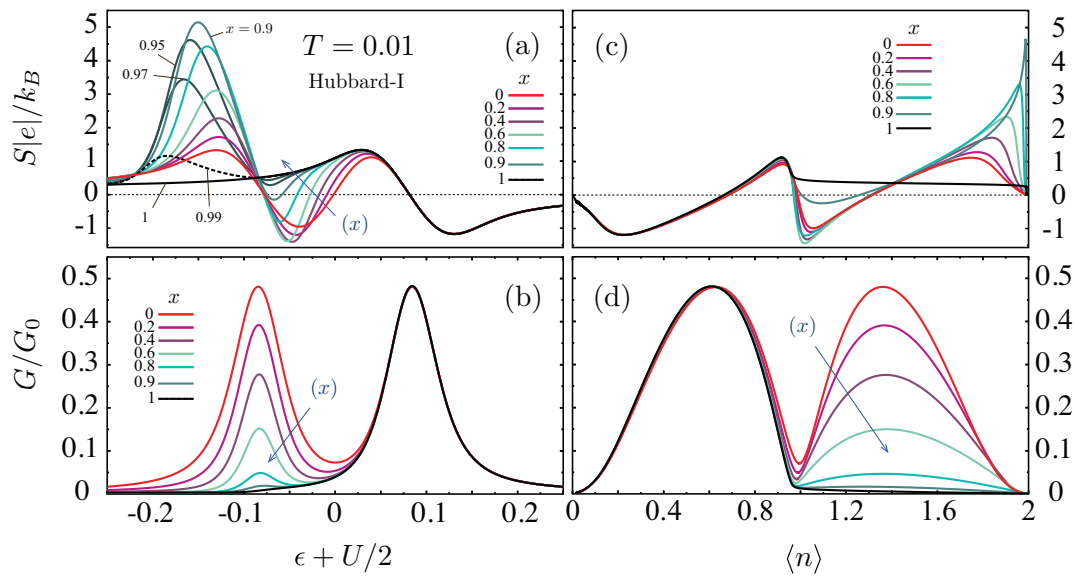}
\caption{\label{fig4}(a) Thermopower $S$, derived from equation~(\ref{S-res}), and (b) conductance $G$, both within the Hubbard-I approximation and as a function of the gate voltage  for different values of assisted hopping $x$, at $T/D=0.01$.
The results overlap with the NRG data (figure~\ref{fig3}) in the empty/full orbital regime. In panels (c) and (d) are shown $S$ and $G$ as a function of $\langle n\rangle$.}
\end{figure}

Results presented in figure~\ref{fig3} correspond to a larger temperature regime, $T>T_K$, for the whole range of $x$ presented. It is therefore not surprising that the Hubbard-I approximation is not only qualitatively but also quantitatively reliable here, as is evident from  figure~\ref{fig4} where are shown results  correspondingin to the NRG counterparts from figure~\ref{fig3}. The main difference between the two approaches is in the Coulomb-blockade valley where the conductance obtained by the NRG is higher (and $S$ lower). One reason could be in the co-tunnelling processes -- see also \cite{Turek} -- absent in the Hubbard-I approximation. Part of the discrepancy originates also in not completely negligible effects of the Kondo peak. Another observation is  a systematic shift of the position of the peaks towards lower $\epsilon + U/2$ values with progressively larger $x \to 1$ for the NRG results compared to the Hubbard-I approach.

Figure~\ref{fig5} presents thermopower and conductance at much lower temperature $T/D=10^{-5}$.
For $x=0$, we recover the usual Kondo effect. At the p-h symmetric point, the estimated $T_K=0.00173 \gg T$ and the system is a Fermi liquid. The average charge $\langle n\rangle$ can then be related with the phase shift due to the Friedel sum rule, and therefore the conductance can be written as
\begin{equation}
\label{eqG}
G=G_0 \sin^2\left(\pi \langle n\rangle /2\right),
\end{equation}
with $G_0=2e^2/h$.
The Sommerfeld expansion for $S$ gives a typical metallic-like dependence  \cite{Costi}
\begin{eqnarray}
\label{SA}
S(T)=-\frac{\pi^2 k^2_{B}T}{3|e|}\frac{1}{A(\omega)}\frac{\partial A(\omega)}{\partial \omega} \Bigr|_{\omega=0},
\end{eqnarray}
showing that the slope of the spectral density at $\epsilon_F$ determines the sign of $S$. Using the Friedel sum rule one can rewrite this formula as \cite{Costi1994}
\begin{eqnarray}
\label{SF}
S(T)=-\frac{\pi\gamma T}{|e|}\cot\left(\pi \langle n\rangle/2\right) ,
\end{eqnarray}
where $\gamma \sim 1/T_K$. {Our NRG calculations confirm this dependence in the regime $\langle n\rangle\approx 1$.
$S$ changes its sign at the p-h symmetric point when $\langle n\rangle$ becomes unity.
For $0<x<1$ the crossing point ({\it i.e.}, $S=0$) is shifted to lower gate voltages, similarly  as the $\langle n\rangle=1$ point in figure~1. Plotting $S$ as a function of $\langle n\rangle$ in panel (c) we see that thermopower becomes zero at $\langle n\rangle=1$ for any $x$. We conclude that in the presence of moderate assisted hopping, equation~\ref{SF} still holds if the prefactor $\gamma$ is suitably modified.

\begin{figure}[t]
\centering
\includegraphics[width=0.9\textwidth,clip]{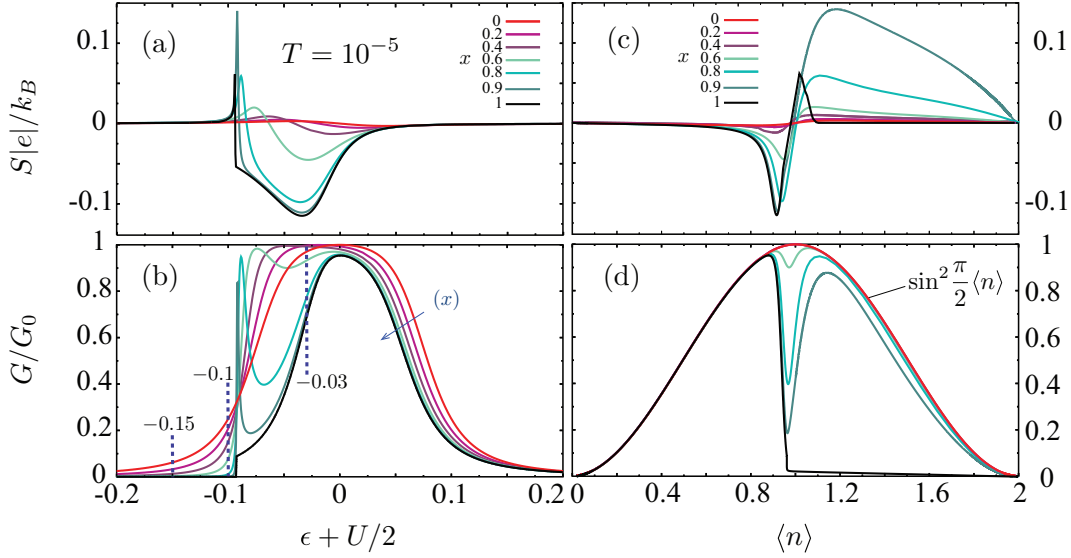}
\caption{\label{fig5} (a) Thermopower $S$ and (b) conductance $G$ as a function of the gate voltage for  $T/D=10^{-5}$. In panel (b) vertical dashed lines indicate special values for $\epsilon+U/2=-0.15$, $-0.10$, and $-0.03$ for which the temperature dependence is in Subsection~\ref{temperature} analysed in detail.
In panels (c) and (d) the same results are plotted
as a function of the occupancy $\langle n\rangle$.}
\end{figure}

Panel (d) shows that the NRG method gives for $x=0$ a perfect shape $G=G_0 \sin^2\left(\pi \langle n \rangle /2\right)$. Assisted hopping, $x\neq 0$, makes the $G$ vs. $\expv{n}$ dependence asymmetrical, reducing the part around $\langle n\rangle =1$. This drop in $G$ is related to a reduction of the exchange coupling $J_K$ (see Section~2.2), which leads to an exponential reduction of $T_K$ below the physical temperature $T$ ({\it i.e.}, we go from the Kondo regime to the Coulomb blockade regime).
Using equation~(\ref{JK1}) for $J_K$, one can express $T_K$ as
\begin{equation}\label{tk}
k_BT_K= \sqrt{U\Gamma/2}\;\exp\{\pi \epsilon(\epsilon+U)/[2\Gamma(U+\epsilon\, x(x-2))]\}.
\end{equation}
We neglect the $x$-dependence of the hybridization in the prefactor as the exponent is much more important. At $x=0$ this formula reduces to the Haldane's expression \cite{haldane1978} $k_BT_K=\sqrt{U\Gamma/2}\; \exp[\pi \epsilon(\epsilon+U)/(2\Gamma U)]$. The minimal value for $J_K$ is  reached at $\epsilon=-U/({1+|1-x|})$ [equation~(6)] and
therefore the minimal value for $T_K$ is given by
\begin{eqnarray}
\label{tkm}
k_BT_K^{min}= \sqrt{U\Gamma/2}\;\exp\{-\pi U/[2\Gamma({1+|1-x|})^2]\}.
\end{eqnarray}
The formulae (\ref{tk})-(\ref{tkm}) show that $T_K$ can be reduced by many orders of magnitude, especially for $x\rightarrow 1$ and  $\epsilon+U/2<0$; it drops below $T=10^{-5}$, leading to a reduction of $G$ and an increase of $S$ in the range $\epsilon+U/2<0$. When $T$ is not sufficiently lower then $T_K$, the Friedel sum rule can
no longer be used to compute $G$ and $S$. Note also that  the conductance plot, figure~\ref{fig5}(b) and (d), qualitatively well reflects the corresponding dependence of the charge fluctuations $\langle \delta n^2 \rangle$, figure~\ref{fig1}(e) and (h), and that both quantities are nearly symmetric with respect to $n_c$ when plotted vs. $\langle n \rangle$.

\subsection{Temperature dependence}
\label{temperature}

\begin{figure}[t]
\centering
\includegraphics[width=0.9\textwidth,clip]{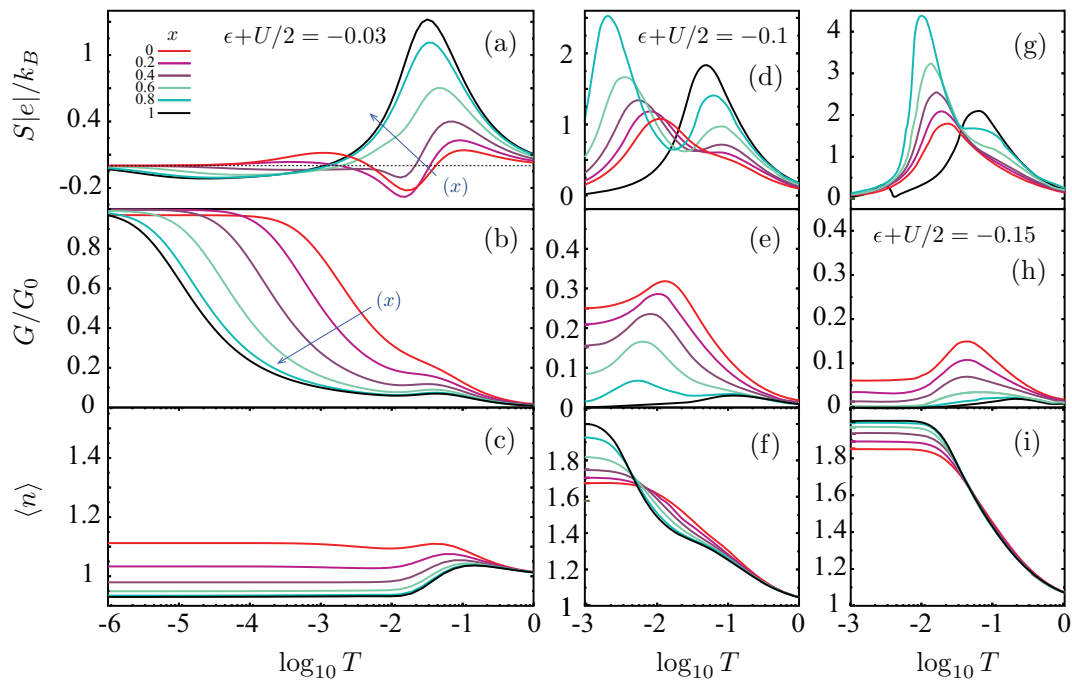}
\caption{\label{fig6} Thermopower $S$, conductance $G$, and occupancy $\langle n \rangle$ as a function of temperature for different values of assisted hopping $x$ and different values of the gate voltage $\epsilon$: left panels corresponds to the Kondo regime ($\epsilon+U/2=-0.03$), middle
bottom panels to valence-fluctuation (between single and double occupancy) at $\epsilon+U/2=-0.1$
and full-orbital regime, $\epsilon+U/2=-0.15$. Positions of these regimes are in figure~\ref{fig5}(b) indicated by vertical  dashed lines.}
\end{figure}

The transport properties and the QD occupancy are shown in figure~\ref{fig6} as a function of $T$, separately for the Kondo, mixed-valence and full-orbital regimes, each having a characteristic temperature dependence. \footnote{In the plots of thermopower and conductance versus gate-voltage in Figs.~\ref{fig3} and \ref{fig5} we had observed that the effect of assisted hopping is more significant for $\epsilon+U/2<0$ compared to $\epsilon+U/2>0$, thus we here focus on the former case.}

The Seebeck coefficient and the conductance are strongly related to the level occupancy, which is also temperature-dependent, $n(T)=\langle n \rangle$. From panels (c),(f),(i) it is clear that $n(T)$ varies mostly on
the valence-fluctuation temperature scale, $\Gamma<T<U$. In the Kondo regime, the variation of $n(T)$ versus $T$ is weak, with $n(T)$ always near half filling.
In the MV regime, the behaviour is more complex, since two
states, at $\epsilon$ and $\epsilon+U$ are at play, thus multiple plateaus can be
discerned for some values of $x$. Finally, in the FO regime, there is a simple evolution from the high-$T$ limit to values of $n(T)$ near 2 at low temperatures.

The temperature dependence of $G$ in different regimes is shown in panels (b,e,h).
The small peaks in $G(T)$ at high temperatures in all regimes are due to  transport through the excited state at $\epsilon+U$.
In the Kondo regime, panel (b),  for $0<x<2$
the Kondo effect is fully developed  at low enough temperatures and $G(T)$ tends to a saturated value $G_0$ near the
unitary limit.
Its exponential dependence is clearly seen in $G$ curves in panel (b): for each $x$, $G(T)$ shows the typical Kondo scaling behaviour at small temperatures, however the curves are shifted to increasingly low temperatures as $x$ is increased.
In both MV, panel (e), and FO, panel (h), regimes,
with increasing $x$ the transport collapses for $T<\Gamma$, which is a consequence
of $n(T)$ tending toward values near 2 in the low-$T$ limit. In addition,
the $G(T)$ peaks in the range $\Gamma<T<U$ are diminished due to the averaging over off-resonant transmission.
In these regimes, there is no Kondo resonance and
the spectral weight at $\epsilon_F$ is low at any $T$.

{\it Thermopower in the Kondo regime}.-
For $x>0.4$ in the Kondo regime, $S(T)$ exhibits a single sign change, as opposed  to two sign changes for $x<0.4$, see panel (a).
Therefore, strong assisted hopping
can change the behaviour of $S(T)$, making the Seeback coefficient behave
like in the FO case.
This complicated behaviour of $S$ has a clear interpretation in terms of
the $x$-dependence of the level positions and hybridization renormalizations.
The effects of $x$ on $S(T)$ can also be
understood as arising
from a change in the local level occupancy $n(T)$.

In the low-temperature Fermi-liquid regime for $T \ll T_{K}$, the behaviour of $G(T)$ and $S(T)$ can be accounted for by considering the structure of the spectral function $A(\omega,T)$ and using the Sommerfeld expansion, Eq.~\eqref{SF}, which
shows that the slope of the spectral density at $\epsilon_F$ determines the sign of $S$. For $\epsilon+U/2<0$, the Kondo resonance lies below $\epsilon_F$ so the slope at $\epsilon_F$ is negative, resulting in a positive thermopower. Similar to semi-metals and metals, the transport occurs near $\epsilon_F$ and $S(T) \propto T$ \cite{Jonson,Costi}. At $T \gtrsim T_{K}$, the sign of $S$ is no longer determined solely by the slope of the spectral function at $\omega=0$, but by the number of states available below and above $\epsilon_F$ in the energy window of order $T$. For $T \gg T_{K}$, similar to semiconductors at low doping conductance where transport only occurs far away from $\epsilon_F$, the system is effectively in the hopping regime in which $S(T) \propto \Delta/T$ \cite{Mott,Seeger}, where $\Delta$ is some characteristic energy for hopping processes.

It should be noted that the thermopower in the Kondo regime does not exhibit universal scaling behaviour, {\it {\it i.e.}}, is not a function solely of $T/T_{K}$. In the Kondo regime it is crucially determined by the potential scattering term which breaks the p-h symmetry and leads to finite value of the Seebeck coefficient. In fact, the conductance is likewise not fully universal in the sense of being solely a function of $T/T_K$. Since the potential scattering in the SIAM is a marginal operator (in the renormalization-group sense), the conductance and thermopower are both also functions of the quasiparticle scattering phase shift. Nevertheless, this dependence is weak for the conductance and mainly manifests as a small deviation of the saturated $T=0$ conductance from the unitary limit, while for thermopower it is a first-order effect which determines the overall scale of
the Seebeck coefficient.

{\it Thermopower for mixed-valence and full-orbital regimes}.-
In the MV regime there are two distinct
peaks in $S(T)$: one associated with the lower level $\epsilon$ below $\epsilon_F$ and
another with the higher level $\epsilon+U$ above $\epsilon_F$.
On increasing $x$, the peaks become sharper due to the renormalization of the levels [see equation~\eqref{epsilon1}].
It should be noted that the peak at low temperatures corresponds to the hole-like transport, while that at high temperatures to electron-like transport.
For $x=1$, there is a single peak, because transport through the state $\epsilon+U$ is not allowed.
The same behaviour is found in the FO regime in which both levels are below $\epsilon_F$
and the hole transport strongly dominates.
Therefore, there is a single peak for FO regime for all values of $x$.

The behaviour of $S(T)$ in the MV regime, panel (d), has been verified using a model of two resonant transmissions at $\epsilon$ and $\epsilon+U$,
and has been found to be in good agreement with the accurate NRG results.
A similar test has been performed for the EO regime, panel (g), where both levels are below $\epsilon_F$.

\section{Conclusions}
\label{conc}

We have investigated the thermopower $S(T)$ and the conductance $G(T)$ of quantum dots described by the Anderson impurity model with assisted-hopping terms. This model takes into account the hybridization processes which depend on the charge state of the impurity; these are present
in real devices, but commonly neglected in simplified theoretical models.
The assisted hopping modifies $G(T)$ and $S(T)$ due to renormalization of the impurity level; in particular, it changes its effective width, but also its position. One of our main findings regarding the thermopower is that, quite generally, there is no sign change in $S(T)$ outside the Kondo regime, making such a sign change in thermopower a particularly sensitive signature of strong correlations and Kondo physics. In the Kondo regime, the behaviour of $G(T)$ is not qualitatively affected by the assisted hopping, only the Kondo temperature scale is strongly modified. The thermopower, however, is much more sensitive to the assisted hopping, because of the different effect of the assisted hopping on the atomic peak in the spectral function at $\epsilon+U$ as compared to that at $\epsilon$, leading to more pronounced
particle-hole asymmetry which the thermopower measures. We thus conclude that the thermopower can be considered as a very sensitive tool to detect assisted hopping and Kondo correlations experimentally.

\ack{Financial support from the EU FP7 project: Marie Curie ITN NanoCTM, the Slovenian Research Agency under contract no. P1-0044, and National Science Centre (Poland) under the contract DEC-2012/05/B/ST3/03208 are gratefully acknowledged. We are also grateful for helpful discussion with Piotr Stefa\' nski.  }

\vskip 1 cm

\end{document}